\documentclass[reprint,
 amsmath,amssymb,
 aps,
floatfix,
showkeys
]{revtex4-1}

\usepackage[USenglish]{babel}
\usepackage{slashed}
\usepackage{physics}
\usepackage{graphicx}% Include figure files
\usepackage{epstopdf}
\usepackage{graphics}
\usepackage{dcolumn}% Align table columns on decimal point
\usepackage{bm}% bold math
\usepackage[usenames]{color}
\usepackage[svgnames]{xcolor}
\usepackage{comment}
\usepackage[colorlinks,citecolor=DarkGreen,linkcolor=DarkRed,urlcolor=DarkBlue]{hyperref}% add hypertext capabilities
\usepackage{float}

\newcommand{\be}{\begin{equation}}
\newcommand{\ee}{\end{equation}}
\newcommand{\bea}{\begin{eqnarray}}
\newcommand{\eea}{\end{eqnarray}}
\newcommand{\beas}{\begin{eqnarray*}}
\newcommand{\eeas}{\end{eqnarray*}}

\begin{document}

\title{Influence of magnetic field-induced anisotropic gluon pressure during
pre-equilibrium in heavy-ion collisions: A faster road towards isotropization}% Force line breaks with \\


\author{Alejandro Ayala$^{1}$, Ana Julia Mizher$^{2,3,4}$}
\affiliation{%
$^1$Instituto de Ciencias Nucleares, Universidad Nacional Aut\'onoma de M\'exico, Apartado Postal 70-543, CdMx 04510, Mexico.\\
$^2$ Instituto de F\' isica Te\'orica, Universidade Estadual Paulista, Rua Dr. Bento Teobaldo Ferraz, 271 - Bloco II, 01140-070 S\~ao Paulo, SP, Brazil.\\
$^3$Laboratório de Física Teórica e Computacional, Universidade Cidade de São Paulo, 01506-000, São Paulo, Brazil.\\
$^4$Centro de Ciencias Exactas and Departamento de Ciencias B\'asicas, Facultad de Ciencias, Universidad del B\'io-B\'io, Casilla 447, Chill\'an, Chile.}
%$^{10}$Centro de Investigaci\'on y Desarrollo en Ciencias Aeroespaciales (CIDCA), Academia Politécnica Aeronáutica, Fuerza A\'erea de Chile, Casilla 8020744, Santiago, Chile.}%


\begin{abstract}

Magnetic fields of a large intensity can be generated in peripheral high-energy heavy-ion collisions. Although their intensity drops fast and, moreover, it is not clear whether these fields last long enough to induce a magnetization during the quark-gluon plasma phase, most of the models and simulations predict a significant intensity that lasts up to proper times of order 1 fm after the beginning of the reaction, which is a typical time for the hydrodynamical phase to start. This interval of time is referred to as the pre-equilibrium stage. The evolution of the reaction during pre-equilibrium is thus likely to be influenced by these fields. In this work we adopt a strong field approximation to study the effects of the magnetic field-induced anisotropy on the gluon pressure. We include this anisotropy within the description obtained by means of Effective Kinetic Theory and explore the consequences to reach isotropization at proper times of order 1 fm. We show that when including the magnetic field effects, isotropization is achieved faster.

\end{abstract}

%\pacs{}% PACS, the Physics and Astronomy
%\keywords{Quantum Chromodynamics, Linear Sigma Model with Quarks, Magnetic Fields}%Use showkeys class option if keyword

\maketitle

%%%%%%%%%%%%%%%%%%%%%%%%%%%%%%%%%%%%%%%%%%%%%%%%%%%%%%%%%%%%%%%%%
\section{Introduction}\label{sec1}

Relativistic heavy-ion collisions are a prime tool to unveil the properties of strongly interacting matter under extreme conditions. Information from the different stages of the reaction can be obtained looking at particular probes that are sensitive to the given conditions during the evolution of the system. For instance, it has been recognized that hard probes are influenced by the dense and hot parton system that dominates at intermediate times of the reaction. During that stage, the use of hydrodynamics has proven to be rather successful. Nevertheless, the description in hydro terms requires that the system reaches a certain degree of thermalization and isotropization, which does not happen instantaneously, but rather, after a time of order $\tau_{\rm hydro}\sim 1$ fm. The evolution during earlier times is usually referred to as the pre-thermal or pre-equilibrium stage.

Pre-equilibrium is characterized by the existence of strong color fields which are liberated after the Glasma, is shattered at the beginning of the reaction. The typical time at which these fields start to dominate the system evolution is of order $\tau\sim 1/Q_s$, where $Q_s\sim 1.4$ GeV is the saturation scale~\cite{iancu2002colour,Gelis:2010nm}. These fields are subject to a strong anisotropic expansion in the longitudinal, or beam direction. The description of the evolution of these saturated gluon fields until they reach the hydro regime, cannot be simply accomplished resorting to classical Yang-Mills theory~\cite{Baier:2000sb,Berges:2013fga,Berges:2013eia}, because isotropization is not reached in such manner when the system is subject to a rapid longitudinal expansion. Instead, the problem has been formulated in the context of Effective Kinetic Theory (EKT)~\cite{Arnold:2002zm,Kurkela:2015qoa, Keegan:2016cpi,Kurkela:2018vqr,Keegan:2016cpi}. The result of the analysis is that, for weak coupling, the gluon occupation number at pre-equilibrium can be given initially in terms of a distribution in momentum space that accounts for an anisotropy coefficient $\xi$ for the longitudinal momentum component, which is taken to be of order $\xi=10$~\cite{Kurkela:2018vqr}. The evolution towards equilibrium with time can be parametrically quantified in terms of the behavior of the ratio of the transverse and longitudinal pressures, $P_T/P_L$ as a function of the gluon occupancy. As time evolves up to $\tau_{\rm hydro}$, $P_T/P_L$ tends to one and the system reaches the gluon occupancy which can then be taken as the initial condition for the hydro evolution. Possible signatures of this pre-equilibrium evolution for final state observables are studied in Refs.~\cite{NunesdaSilva:2020bfs,Garcia-Montero:2024msw,Garcia-Montero:2024lbl,Garcia-Montero:2023lrd,NunesdaSilva:2020bfs}.

In recent times, it has also been realized that other relevant players during pre-equilibrium are the strong electromagnetic fields produced in peripheral collisions~\cite{STAR:2023jdd}. In particular, magnetic fields as intense as $|eB|\sim 10^{19}$ Gauss have been inferred to be present in these systems. The fields are at their peak intensity precisely during the pre-equilibrium stage to subsequently fade down fast. Such intense fields contribute to the QCD effective
potential~\cite{Galilo:2011nh,Nedelko:2022kjy} and can leave their imprints~\cite{Pu:2016bxy}, particularly on penetrating probes that are not much sensitive to the hadronic part of the system evolution. Photons (both real and virtual) are precisely one of those penetrating probes. The presence of magnetic fields allows the opening of channels for photon production, otherwise precluded in the absence of these fields. Two of these channels, which are relevant for photon production, are the gluon fusion and gluon splitting. Since at pre-equilibrium, the gluon occupancy is large and quarks are basically absent~\cite{Garcia-Montero:2019vju,Monnai_2020}, gluon driven process for photon production become even more relevant. Given that the magnetic field provides a preferred direction pointing transverse to the reaction plane, it serves also as a natural source of a positive $v_2$. 

In a series of recent works, the above mentioned processes have been explored, with the focus in the description of the one-loop matrix element for very intense fields~\cite{Ayala:2024ucr,Ayala:2022zhu,Ayala:2019jey,Ayala:2017vex}. However, another aspect of the problem, that to our knowledge has not yet been addressed, is the fact that the magnetic field also produces an anisotropic pressure between the parallel  ($\parallel$) and perpendicular ($\perp$) directions with respect to the magnetic field and thus, its contribution to the overall anisotropy during pre-equilibrium needs to be accounted for to explore its effect for the road towards isotropization. To simplify the problem, in this work we compute the effects of an intense and homogeneous magnetic field, pointing transverse to the reaction plane, on the gluon contribution to the anisotropy. It may be thought that since gluons possess no electric charge, they are unaffected by a magnetic field. However, due to quantum fluctuations involving quarks, gluons feel the presence of the magnetic field and behave as if moving in a magnetized medium, modifying their dispersion properties~\cite{Ayala:2018ina,Ayala:2019akk,Fukushima:2011nu,Hattori:2017xoo}. When the field is intense, only the parallel component of the polarization tensor is present. Since during pre-equilibrium the field is very intense, one can then approximate the gluon dispersion properties as described only by the longitudinal polarization. In this work we use such approximation and compute the anisotropy in the pressure induced by the magnetic field. For simplicity the field is taken as homogeneous and its evolution in time described in terms of the retarded potentials produced by spectators and participants in a heavy-ion collision. Other more sophisticated approaches to describe the field evolution during pre-equilibrium, such as the one studied in Ref.~\cite{Yan:2021zjc}, are also possible. We also explore the consequences when this extra pressure is accounted for to reach isotropization at the end of the pre-equilibrium stage.

Recall that the gluon polarization tensor in the strong field limit, where the quark mass can be neglected, can be written as~\cite{Ayala:2018ina,Ayala:2019akk,Fukushima:2011nu,Hattori:2017xoo}
\be
\Pi^{\mu\nu}=g^2\left(g^{\mu\nu}_\parallel-\frac{q^\mu_\parallel q^\nu_\parallel}{q^2_\parallel}\right)\sum_f\frac{|q_fB|}{8\pi^2}e^{-q_\perp^2/(2|q_fB|)},
\label{poltens}
\ee
where $g$ is the strong coupling. We sum over the light quark species $f=u,d$ with $q_f$ representing their corresponding charge (in units of the electron charge),  and we define the metric components in the longitudinal and transverse (to the magnetic field) direction, as
\begin{eqnarray}
g^{\mu\nu}_\parallel&=&{\mbox{\rm diag}}\ (1,0,0,-1),\nonumber\\
g^{\mu\nu}_\perp&=&{\mbox{\rm diag}}\ (0,-1,-1,0),
\label{metrics}
\end{eqnarray}
such that $g^{\mu\nu}=g^{\mu\nu}_\parallel+g^{\mu\nu}_\perp$. Therefore
\begin{eqnarray}
q^\mu_\parallel&=&g^{\mu\nu}_\parallel q_\nu\nonumber\\
q^\mu_\perp&=&g^{\mu\nu}_\perp q_\nu,
\label{components}
\end{eqnarray}
and $q^2_\perp=-q^\mu_\perp q^\perp_\mu$. Using this expression, the magnetic field dependent gluon propagator in the strong field limit can be written as
\begin{eqnarray}
iG^{\mu\nu}= \frac{\left(g^{\mu\nu}_\parallel-q^\mu_\parallel q^\nu_\parallel/q^2_\parallel\right)}{q_\parallel^2 - q_\perp^2 - g^2\sum\limits_f\frac{|q_fB|}{8\pi^2}e^{-q_\perp^2/(2|q_fB|)}}.
\label{propagator}
\end{eqnarray}

The pressure in the parallel direction can be computed from the effective potential $V$ as $P_\parallel=-V$. At one-loop order, and after performing a Wick rotation to Euclidean space, this is given by
\begin{eqnarray}
V&=&-\frac{i}{2}\int\frac{d^2q_\parallel} {(2\pi)^2}\frac{d^2q_\perp} {(2\pi)^2}\nonumber\\
&\times&\ln\!\left(-q^2 + g^2\sum\limits_f\frac{|q_fB|}{8\pi^2}e^{-q_\perp^2/(2|q_fB|)}\!\right),
\label{effpot1}
\end{eqnarray}
with $q^2=q_\parallel^2-q_\perp^2$.

Notice that modes with $q_\perp^2$ larger than twice the field strength are basically insensitive to the magnetic field. Therefore we can approximate the exponential in Eq.~(\ref{effpot1}) taking the first order in the Taylor expansion to write
\begin{eqnarray}
V&=&-\frac{i}{2}\int\frac{d^2q_\parallel} {(2\pi)^2}\frac{d^2q_\perp} {(2\pi)^2}\nonumber\\
&\times&\ln\!\left(-q_\parallel^2 + \left[1-\frac{g^2}{8\pi^2}\right]q_\perp^2 + g^2\sum\limits_f\frac{|q_fB|}{8\pi^2}\!\right)\nonumber\\
&=&
\frac{i}{2a}\int db\int\frac{d^2q_\parallel} {(2\pi)^2}\frac{d^2\tilde{q}_\perp} {(2\pi)^2}\frac{1}{(q_\parallel^2-\tilde{q}_\perp^2 - b)},
\label{effpot2}
\end{eqnarray}
where we have defined
\begin{eqnarray}
a&\equiv&\left[1-\frac{g^2}{8\pi^2}\right]=\left[1-\frac{\alpha_s}{2\pi}\right]\nonumber\\ b&\equiv&g^2\sum\limits_f\frac{|q_fB|}{8\pi^2}=\frac{\alpha_s}{2\pi}|eB|\nonumber \\
\tilde{q}_{\perp}^2&\equiv&a\ q_\perp^2,
\end{eqnarray}
where we used that $\alpha_s=g^2/4\pi$ and $q_u=2/3\ e$, $|q_d|=1/3\ e$. For the calculation we take $\alpha_s=0.3$

We start by performing the integration in the parallel variables. We foresee that the divergence are associated to motion along these variables since, contrary to the variables in the perpendicular direction, in the large field limit they are not altered by the presence of the magnetic field. The result is
\begin{eqnarray}
V=\frac{1}{(8\pi)a}\int db\int\frac{d^2\tilde{q}_\perp}{(2\pi)^2}\left[\frac{1}{\epsilon}+\ln\left(\frac{\mu^2}{\tilde{q}_\perp^2+b}\right)\right],
\label{effpot3}
\end{eqnarray}
where we worked in $d=2-2\epsilon$ dimensions with $\mu$ the renormalization scale in the $\overline{MS}$ scheme. As anticipated, a divergence, expressed by the $1/\epsilon$ term, appears. This can be absorbed in the renormalization of the coupling $g$. We thus concentrate in the second term of Eq.~(\ref{effpot3}). Recall that this expression is valid for perpendicular momenta $q_\perp^2\leq\Lambda^2\sim 2|eB|_{\max}$. Simulations~\cite{Ayala:2019jey} show that the maximum values of the magnetic field for collision energies of order $\sqrt{s_{NN}}=200$ GeV for mid to periferal centralities are of order $|eB|_{\max}\gtrsim m_\pi^2$.  To cover all cases of interest, we thus take $\Lambda=4 m_\pi$ and perform the integration up to this hard cut off followed by integration over $b$ with the result
\begin{eqnarray}
V&=&\frac{1}{(32\pi^2)a}\int db\left[a\Lambda^2+b\ \ln\left(\frac{b}{\mu^2}\right)\right.\nonumber\\
&+&\left.(b+a\Lambda^2)\ln\left(\frac{\mu^2}{b+a\Lambda^2}\right)\right]\nonumber\\
&=&\frac{1}{(64\pi^2)a}\left[3a\Lambda^2 b+b^2\ln\left(\frac{b}{(\Lambda_{QCD}/2)^2}\right)\right.\nonumber\\
&+&\left.\left(b + a\Lambda^2\right)^2\ln\left(\frac{(\Lambda_{QCD}/2)^2}{b+a\Lambda^2}\right)\right],
\label{effpot4}
\end{eqnarray}
where we have set the renormalization scale $\mu=\Lambda_{QCD}/2=100$ MeV, representing a typical scale of the confining/deconfining transition. 

\begin{figure} [t]
\centering
\includegraphics[width=0.45\textwidth, ]{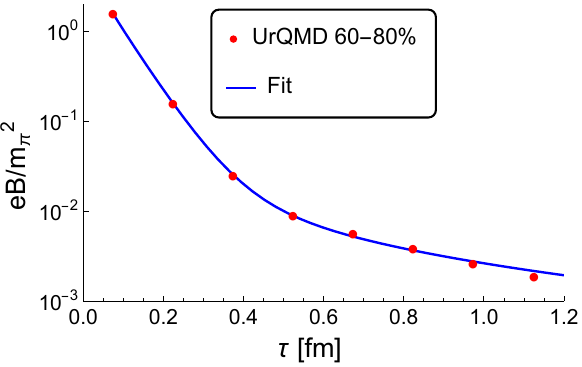}
\caption{Magnetic field strength in units of the pion mass squared as a function of proper time. The dots are the result of a UrQMD simulation considering the contribution of both the participants and the spectators in Au+Au semi-central collisions, 60-80\% centrality, at $\sqrt{s_{NN}}=200$ GeV. The continuous curve is a fit (see text) to the UrQMD simulation.}
\label{fig1}
\end{figure}
For systems where the electric conductivity is infinite, the perpendicular pressure can be computed as $P_\perp=P_\parallel- M|eB|$, where the magnetization $M$ is given by $M=-\partial V/\partial |eB|$~\cite{Bali:2014kia}. These systems have the magnetic flux frozen in the medium as they evolve with time. However, although unknown, it is very unlikely that the magnetized gluon medium in pre-equilibrium corresponds to one with infinite conductivity. Recall that the electrical conductivity is a measure  of
how easily the electric charge carriers flow in response to an applied electric field. In pre-equilibrium, although few, these charge carriers are the quarks. Since quarks follow trajectories around the field lines, the motion in the perpendicular directions is constrained,  resulting in a significant suppression of
the transport in these directions with respect to the parallel direction~\cite{Hattori:2016cnt,Ghosh:2024fkg}.
\begin{figure} [t]
\centering
\includegraphics[width=0.45\textwidth, ]{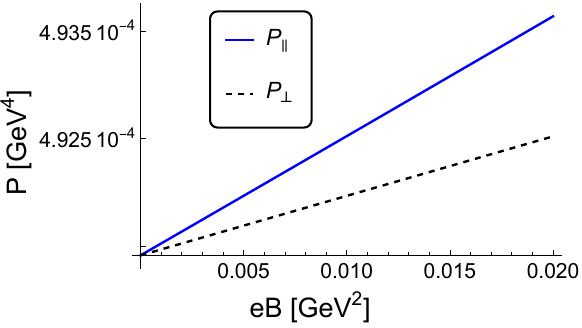}
\caption{Longitudinal and perpendicular magnetic field induced pressures as functions of the field strength. $P_\parallel$ is shown with the solid line. $P_\perp$ is shown with the dashed line.}
\label{fig2}
\end{figure}
\begin{figure}[b]
\centering
\includegraphics[width=0.45\textwidth, ]{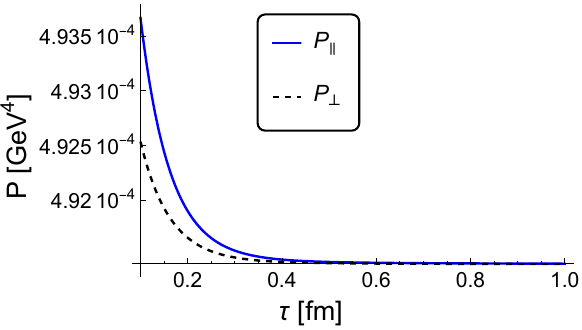}
\caption{Parallel and perpendicular magnetic field induced pressures as functions of the proper time. $P_\parallel$ is shown with the solid line. $P_\perp$ is shown with the dashed line.}
\label{fig3}
\end{figure}
For example, in a static QGP medium at finite temperature and baryon density,  almost constant ratios of the conductivity to the temperature, of order one tenth, have been reported~\cite{Thakur:2019bnf}. For systems where the magnetic field profile has a spatial dependence, the compression
perpendicular to the magnetic field depends on the trajectory $x_i$ such that $P_\perp=P_\parallel+Mx_i\partial|eB|\partial x_i$.
To obtain a quantitative estimate, without invoking any particular spatial profile, let us consider the expression for the pressure in the perpendicular direction given by
\begin{eqnarray}
P_\perp=P_\parallel - \eta M|eB|,
\label{transverse}
\end{eqnarray}
and take $\eta= 0.5$.

Figure~\ref{fig1} shows the magnetic field strength, in units of the pion mass squared, as a function of proper time, produced both by the participants and spectators in Au+Au collisions at $\sqrt{s_{NN}}=200$ GeV for semi-central collisions with 60-80\% centrality, as found in Ref.~\cite{Ayala:2019jey}. The time dependence of the field strength can be parametrized as
\begin{eqnarray}
\frac{|eB|}{m_\pi^2}= Ae^{-B\tau}+\frac{C}{\tau^D},
\label{profile}
\end{eqnarray}
with $A=4.432$, $B=15.895$ fm$^{-1}$, $C=0.003$ fm$^{D}$ and $D=1.682$ with $\tau$ given in fm.

Figure~\ref{fig2} shows the longitudinal and perpendicular magnetic field induced pressures, obtained from Eqs.~(\ref{effpot4}) and~(\ref{transverse}), as functions of the field strength. Notice that, as expected, $P_\perp < P_\parallel$. 

Figure~\ref{fig3}
shows the evolution of the parallel and perpendicular pressures with proper time. We start the evolution at  $\tau_0=0.1$ fm and stop it at $\tau_f=1$ fm. As expected $P_\parallel$ remains larger than $P_\perp$ throughout of the entire time evolution, being about 25\% larger at the beginning of the time evolution.

\begin{figure}[t]
\centering
\includegraphics[width=0.45\textwidth, ]{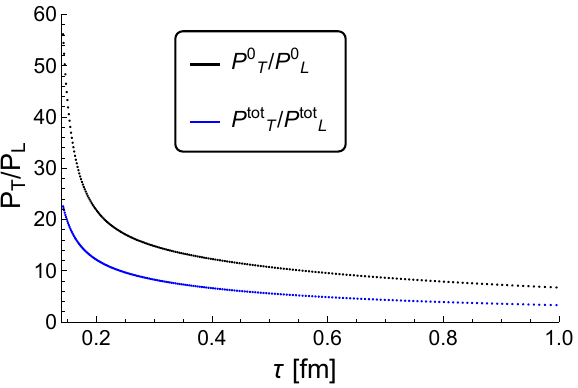}
\caption{Comparison between the time evolution of the ratio $P_T/P_L$ with and without  the magnetic field induced contribution.}
\label{fig4}
\end{figure}

The parallel and perpendicular pressures contribute during pre-equillibrium to the transverse and longitudinal (with respect to the beam direction) pressures discussed in Ref.~\cite{Kurkela:2015qoa,Kurkela:2018vqr}, respectively such that
\begin{eqnarray}
P_T&\to&P_T^{\text tot}=P_T+P_\parallel\nonumber\\
P_L&\to&P_L^{\text tot}=P_L+P_\perp,
\label{newpressures}
\end{eqnarray}
and therefore influence the evolution towards isotropization. Figure~\ref{fig4} shows evolution of $P_T^{\text tot}/P_L^{\text tot}$ as a function of time. For comparison, we also show the ratio $P_T^0/P_L^0$, namely, without considering the magnetic field induced pressures, found in Ref.~\cite{Kurkela:2015qoa}, computed with a t'Hooft coupling $\lambda=10$ and an initial anisotropy parameter $\xi=10$. 
Notice that when the magnetic field induced pressure is considered, isotropization, namely, a value of $P_T/P_L$ closer to 1 is achieved faster. This can be understood from the results illustrated in Figs.~\ref{fig2} and~\ref{fig3} which show that, although $P_\perp < P_\parallel$, they are comparable in magnitude. Since, on the other hand, according to the results of Ref.~\cite{Kurkela:2015qoa}, $P_T>>P_L$ and also since $P_T\sim P_\parallel\sim P_\perp$, when forming the ratio $P_T^{\text tot}/P_L^{\text tot}$, this behaves as
\begin{eqnarray}
\frac{P_T^{\text tot}}{P_L^{\text tot}}\sim\frac{P_T+P_\parallel}{P_\perp},
\label{ratiooftots}
\end{eqnarray}
and thus this ratio becomes naturally smaller than $P_T^0/P_L^0$, because in the denominator of Eq.~(\ref{ratiooftots}) the small quantity $P_L$ has been replaced by the larger quantity $P_\perp$.

\begin{figure}[t]
\centering
\includegraphics[width=0.45\textwidth, ]{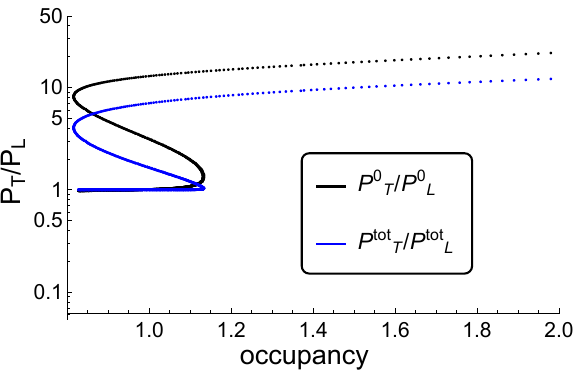}
\caption{Comparison between the ratios $P_T/P_L$, with 
and without the magnetic field induced contribution, as functions of the occupancy.}
\label{fig5}
\end{figure}
%Finally, Fig.~\ref{fig5} shows the ratio of the transverse to the longitudinal pressures, as functions of the occupancy, defined as the average of the gluon distribution times the gluon momentum divided by the average of the gluon momentum $\langle fp\rangle/\langle p\rangle$, with and without accounting for the contribution of the magnetic field induced pressures. Notice that when the magnetic field induced pressure is considered, the ratio of pressures $P_T^{\text tot}/P_L^{\text tot}$ becomes flat as a function of occupancy and that its value is close to 1 during the entire evolution.

Finally, Fig.~\ref{fig5} shows the ratio of the transverse to the longitudinal pressures, as functions of the occupancy, defined as the average of the gluon distribution times the gluon momentum divided by the average of the gluon momentum $\langle fp\rangle/\langle p\rangle$, with and without accounting for the contribution of the magnetic field induced pressures. Notice that when the magnetic field induced pressure is considered, the ratio of pressures $P_T^{\text tot}/P_L^{\text tot}$ is smaller in all the stages of the evolution and the value of the occupancy where this ratio reaches 1 is slightly larger than the case where no magnetic field effects are considered. We emphasize that for the analysis, we employ the results obtained in Ref.~\cite{Kurkela:2015qoa}, where the occupancy is computed without magnetic field induced effects. A full analysis should consider the magnetic field effects already in the EKT to account for the dependence of $\langle fp\rangle/\langle p\rangle$ on time, which is therefore  also likely to change.

To summarize, we have studied the effects of the anisotropic gluon pressure induced by the presence of a strong magnetic field during the pre-equilibrium stage of the evolution of a peripheral high-energy heavy-ion collision. In the approximation where the magnetic field is strong, we have shown that the parallel (to the field direction) induced pressure is larger than the perpendicular pressure albeit both are comparable in magnitude. When adding these pressures to the transverse and longitudinal (with respect to the beam direction, respectively), in the absence of the field, we have shown that isotropization during pre-equilibrium is achieved faster. To gain analytical insight into the magnetic field effects on the evolution of the balance between transverse and longitudinal pressures, the calculation resorted to the strong field limit, whereby only the parallel component, with respect to the field direction, of the gluon polarization tensor contributes. As such, the results of this work should be considered as approximate. A more complete calculation requires relaxing the strong field approximation to consider the contribution of the three possible polarization states to the gluon propagator, and thus to the magnetic field induced pressure. In addition, the effects of the magnetic field induced anisotropy between the perpendicular direction to the beam in the reaction plane and the direction of the magnetic field, requires being explored. These calculations are currently being performed and will be soon reported elsewhere as part of a more extensive work.

\section*{Acknowledgements}

The authors thank  A. Kurkela and F. Lindenbauer for their kind help preparing and sharing their data in tabular form, J.D. Casta\~no-Yepes, I. Dom\'\i nguez, J. Salinas, and M.E. Tejeda-Yeomans for preparing the data of the time evolution of the magnetic field in tabular form and J.J. Medina for his kind help with the handling of the data to produce the plots. A.A. is in debt to the IFT-UNESP and Unicid for their kind hospitality during the time this work was conceived. Support for this work was received in part by UNAM-PAPIIT grant number IG100322 and by Consejo Nacional de Humanidades, Ciencia y Tecnolog\'ia grant number CF-2023-G-433. Support for this work was received in part by grant 2023/08826-7 from the São Paulo Research Foundation (FAPESP). 

\nocite{apsre.v41Control}
\bibliographystyle{apsrev4-1}

\bibliography{biblio,revtex-custom}

\end{document}